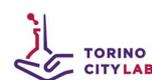
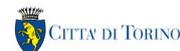
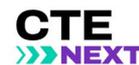

# Literature Survey on how to cluster and define Living Labs, Real World Laboratories and similar research infrastructures

## Authors

Troung Giang Luu[1], Tanja Zylowski[2], Sascha Alpers[2], Andreas Oberweis[1,2]
1 Karlsruher Institut für Technologie, Karlsruhe, Germany
2 FZI Forschungszentrum Informatik, Karlsruhe, Germany

## Abstract

In today's world, where societal challenges in the areas of digitalization, demographic change and sustainability are becoming increasingly complex, new innovation structures are needed to meet these challenges. Living Labs or also Real World Laboratories prove to be such. Through their applied methods such as co-creation, they integrate users into research, making it more user-centric. Which other research infrastructures exist and how they can be differentiated is presented in this paper on the basis of a systematic literature research. Furthermore, methods for user integration are examined and provided in the form of an overview.

## 1. Introduction

The systematic development of innovations is important to counteract the challenges and problems, socially as well as ecologically, which can have a lasting impact on the future of society. Pollution of the environment and climate by new buildings, increasing world population and associated problems of mobility in urban areas are just a few of many examples (Rat für Nachhaltige Entwicklung - Council for Sustainable Development, 2022). Innovations can be explored in many different areas, such as energy, media, mobility, or healthcare (Schäpke et al., 2017). Car-to-car communication in the field of mobility is one such innovation (Becker, 2015). In order to successfully bring such innovations to the market, a suitable environment is needed to realize and test these ideas. In order to increase the success and market acceptance of innovations, affected user groups and other stakeholders are increasingly involved in the research. These are often researched with the help of certain research infrastructures. According to the definition of the European Strategy Forum on Research Infrastructures (ESFRI), a research infrastructure is a unique facility, resource, and service, publicly or privately owned, that is established specifically for scientific purposes, tends to be provided on a medium-term to permanent basis, and requires specific scientific or interdisciplinary (methodological) competencies for its proper establishment, operation, and use (European Commission, 2015). The function of a research infrastructure is to facilitate and enable research and teaching (Wissenschaftsrat, 2012). In the wake of growing societal challenges, existing research infrastructures have been adapted and new ones are created. Which types of them are currently discussed in research and in which areas they are applicated will be investigated in this paper. With the help of a systematic literature review, a detailed overview of novel participatory research infrastructures and their characteristics is provided.

## 2. Novel research infrastructures

In the course of increasingly complex societal challenges, new research infrastructures have been developed that make it possible to meet these new societal challenges. These infrastructures focus on integrating users into research and therefore making development user-centered. The European Commission states that open science with new types of research infrastructures can meet the new societal challenges and make research and innovation more efficient, more creative and stronger (European Commission, 2020). The European Commission has been working with member states and scientific communities to develop new research infrastructures and expand existing infrastructures through the European Strategy Forum on Research Infrastructures (ESFRI) (European Commission, 2020). However, a precise definition of this new type of research infrastructures is not provided by the European Commission. In a publication provided by the German Bundestag, an approach for the description of such new research infrastructures can be found. A novel research infrastructure is described as a new type of cooperation between science and civil society that enables mutual learning in an experimental environment (Deutscher Bundestag, 2018). Thus the active participation of the population in scientific research by providing intellectual collaboration, local knowledge or provision of funds and resources is required (Deutscher Bundestag, 2018). Societal participation, for example with the use of co-creation, as well as participating stakeholders in research is therefore considered a central element of novel research infrastructures (Rose et al., 2019).  Consequently, in this work a novel research infrastructure contains the following characteristics:
▪ Cooperation between researchers and other civil society actors
▪ Learning and exploration in an experimental setting
▪ Active participation of actors in research, and
▪ User-centered research development.

User-centered research development, in particular, is of great importance in novel research infrastructures. From the research perspective, this helps to gather feedback on work, ideas, and concepts that are being developed to generate new forms of knowledge.

## 3. Research Method

The systematics of the applied literature research is based on Kitchenham's method in order to enable a comprehensive and comprehensible review and reporting (Kitchenham, 2007). In the first step, a rough literature search on the topic was conducted in order to become familiar with the topic on the one hand and to get an approximate overview of the amount of existing literature on the other hand. Only then was the design of the systematic literature search developed. For this purpose, first the search strategy was defined. For the search strategy, the research questions were broken down into individual criteria. This allowed a list of synonyms, abbreviations, and alternative spellings to be generated, which formed the basis for the search string (Table 1). The PICOC (Population, Intervention, Comparison, Outcomes, Context) criteria according to Petticrew and Roberts (2008) were used for the decomposition. PICOC is a method that allows research questions to be narrowed into categories using specific criteria. For example, the Comparison criterion questions what the intervention is being compared to or the context in which the intervention is being conducted using the Context criterion. By using the PICOC criteria, it can be ensured that relevant aspects of the research questions are considered in the creation of the search string. For this work, the criteria are defined as follows:

▪ Population: collaboration with stakeholders or users integrated into the research
▪ Intervention: cocreation
▪ Comparison: distributed or concurrent development, process-oriented
▪ Outcomes: reduced cost, improved research quality
▪ Context: novel research infrastructure.

To include all relevant studies, the Context criterion was applied first. While reading into the topic and matching the definition of a novel research infrastructure in this work, three research infrastructures were already potentially identified (Living Labs, Real World Laboratories, test beds). Thus, the terms for the already known three novel research infrastructures (Living Labs, Real World Laboratories, test beds) were added to the search string. The search string was only expanded with additional terms from the PICOC criteria when the number of results from the entered search string was too high to evaluate. Due to the time frame of the present work, this limit was set to 1,000 hits. Additional terms could be added by looking at keywords from already identified publications and databases. Through this consideration, the term "Innovation Infrastructure" was frequently discovered in the context of Living Labs, Real World Laboratories and test beds. Therefore, the term "Innovation Infrastructure" was added to the search string with the assumption of finding additional potential novel research infrastructures. In addition to the full systematic review, reference lists of relevant primary studies and review articles were added to the search. To conduct a comprehensive search, eight different Internet-based databases (EBSCO Business Source Premier, EBSCO Academic Search, IEEEXplore, ProQuest ABI Inform, ACM Digital Libary, ScienceDirect, Google Scholar, dblp) were identified that were relevant to this study. Boolean operators AND and OR were used to link the complex search strings from the defined search terms. The literature search was concluded on 01/31/2022. The search strings used are shown in Table 1.

**Table 1.** Defined search strings

| Criterion | Criterion | Search string |
|---|---|---|
| Basis search string | Novel Research Infrastructure | Innovative research infrastructure OR neuartige Forschungsinfrastruktur OR partizipative Forschungsinfrastruktur OR Innovation Laboratory |
| | Living Labs | Living Labs |
| | Real World Laboratory | Reallabor OR Real-World Laboratory |
| | Test bed | Testfeld OR Testfield |
| AND | | |
| Additional restrictions | Collaboration | Kollaboration OR Collaboration OR Kooperation OR User Integration OR Nutzerintegration OR User Centered |
| | Intervention | Kokreation OR Ko-creation OR Cocreation |
| | Comparison | Concurrent engineering OR Simultaneous engineering OR verteilte Entwicklung OR gleichzeitige Entwicklung |

Exclusion criteria were defined for the review of the sources in order to make the relevance assessment of the articles transparent and reproducible. A total of five exclusion criteria were defined: language, duplicates, publication year, and out of topic. During the review of the sources, the predefined exclusion criteria were applied. This was done in stages by triage of the titles and triage of the abstract. Literature sources that could not be excluded through this triage were further considered for the full text search. These were then carefully read through to determine relevance to the research question.

**Table 2.** Exclusion criteria

| Criterion | Description |
|---|---|
| Language | All sources written in a language other than English or German are excluded. |
| Duplicates | Sources already in our inventory or added by our search strategy are also excluded to avoid duplicates . |
| Publication year | The first novel research infrastructures were discussed scientifically in the early 21st century - for example, Living Labs (M. Eriksson et al. 2006). Therefore, the search is limited with respect to the date of publication in order to obtain relevant results for our study. Sources before 1990 are therefore excluded. |
| Out of topic | Research that by definition does not belong to a novel research infrastructure is excluded. Furthermore, research that does not include methods or characteristics of research infrastructures is excluded. |

The results of the literature search are shown in form of a flow diagram in Figure 1.

**Figure 1:** Results of the literature search

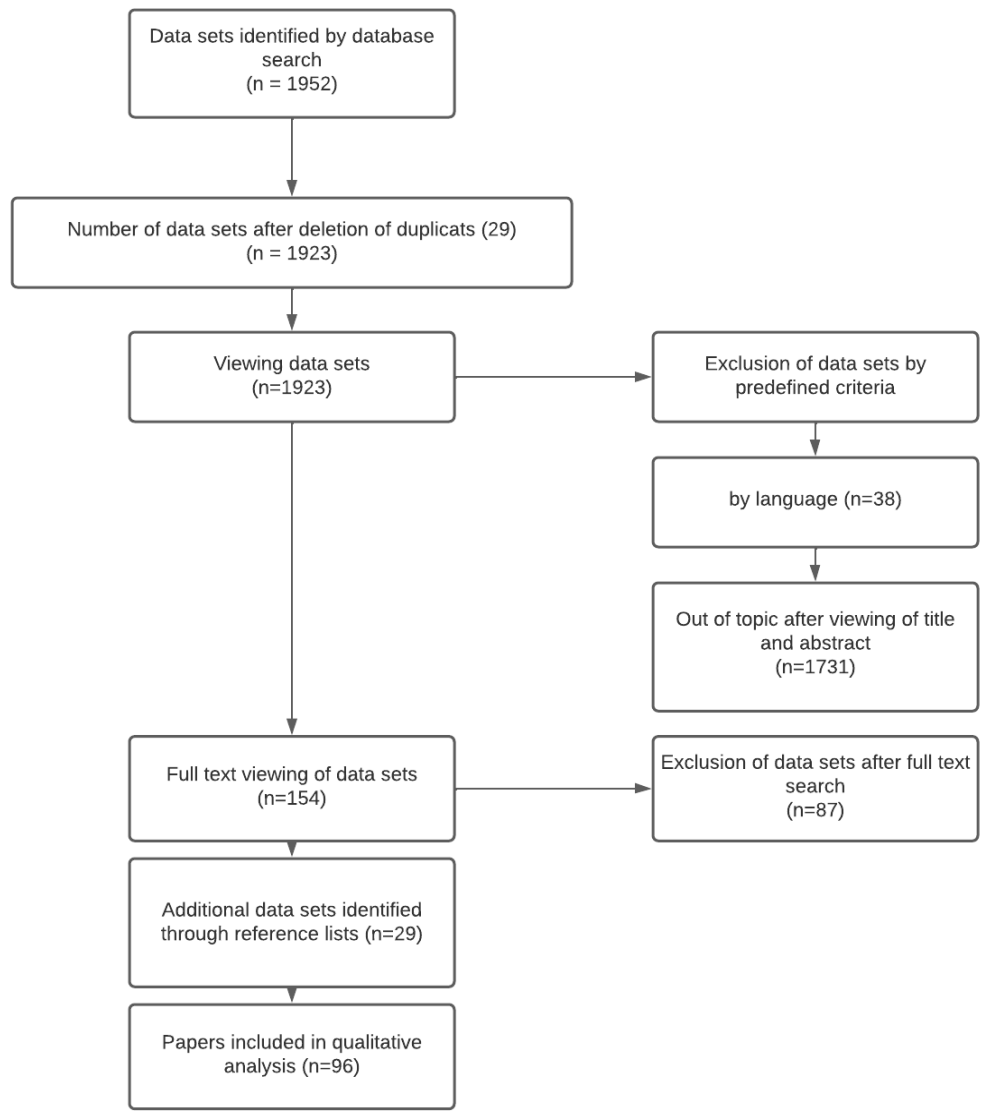

A total of 1952 potentially relevant sources were identified through the defined search strings from the eight different databases. After removing duplicates, applying exclusion criteria, and adding relevant sources through reference lists, a total of 96 sources were abstracted for qualitative analysis. For the EBSCO Business Source Premier, EBSCO Academic Search, IEEEXplore, ProQuest ABI Inform, ACM Digital Libary, ScienceDirect, and dblp databases, the procedure in Table 1 was straightforward. Searching with Google Scholar had to be done in a separate way. Since the number of hits using the basic search string was 18300 search results and it was not possible to achieve results below 1000 with the defined further restrictions from Table 1, an alternative procedure for limiting the number of sources was carried out. The basic search string is still used. Additionally, the search string ["systematic literature review" OR "literature review" OR "literature reviews] was combined with the AND operator. As a result, only research papers that conducted a systematic or simple literature review related to the terms in the base search string were included in Google Scholar. For the purpose of standardization, this particular search string was also additionally performed in the EBSCO Business Source Premier, EBSCO Academic Search, IEEEXplore, ProQuest ABI Inform, ACM Digital Libary, ScienceDirect, and DBLP databases.

To analyse the sources considered relevant, the thematic analysis by Braun and Clarke (2006) was applied. With the help of this thematic analysis, the characteristics of the identified research infrastructures from different sources can be analysed and a typification can be designed. In applying thematic analysis, the following six phases were followed: Phase 1: Initial Data Review. In this step, the entire collected dataset is read through to become familiar with it. This helps to form patterns and ideas prior to coding. Additionally, the depth and breadth of coding can be assessed here. Phase 2: Initial Coding Generation. In this phase, initial coding is generated. These describe characteristics of the data and relate to basic elements of the raw data. Phase 3: Topic Search. In this phase, all previously coded data is sorted and classified into potential themes. After this phase, one has a collection of candidate themes and sub-themes and all data extracts coded in relation to these themes. Phase 4: Topic Review. Here the themes are reviewed again and further refined. Candidate themes may be discarded if there is not enough data to support them or if the data is too diverse. Furthermore, topics that overlap may be merged here while other topics may need to be split into separate topics. Phase 5: Defining and naming the themes. This is where aspects of the data from each theme are captured. This allows them to be better refined and defined. Phase 6: Creation of the report. With the topics fully present and developed, the final phase is to write the report and final analysis.

## 4. Results

A total of 96 relevant publications were identified in the systematic literature search. Out of this 96 publications 91 are peer-reviewed, 1 is an article from a professional journal, 1 is a scientific working paper, 1 is a scientific report and 2 are master theses. Based on the 96 publications found, three different novel research infrastructures were defined. The identification of the three novel research infrastructures is based on the definition of novel research infrastructures given above. Accordingly, the defined criteria ▪ Cooperation between scientists and other actors from civil society ▪ Learning and exploring in an experimental environment ▪ Active participation of actors in research and ▪ User-centered research development were matched with the description of the research infrastructure in the individual publications. In the next step, a typification of the identified novel research infrastructures was created and three types of novel research infrastructures were identified. The following research infrastructures were identified: Living Labs, Real World Laboratories and novel test beds. From the 96 identified publications 48 papers focused on Living Labs, 4 on Real World Laboratories, 3 on novel test beds, and 41 mixed up the defined terms. In order to provide relevant criteria for the distinction of the three concept the most descriptive and unambiguous sources are cited. Sources that were more abstract in their results were analysed but are not cited.

### a. Living Labs

For this work, the Living Lab definitions of Ballon and Schuurman (2015) and the European Network of Living Labs ENoLL (2022) are used as they describe the elements of Living Labs in detail. In summary, therefore, Living Labs are defined in this paper as user-centered, open innovation ecosystems based on a systematic approach that integrates, tests, and experiments research and innovation processes in communities and environments by actively engaging the user and other stakeholders. Open innovation is the targeted opening of the innovation process to the outside world (Liedtke et al., 2012). This makes it possible to involve external parties such as partners, research institutes or even customers in the development process of new ideas, products or services. Thus, Open Innovation is related to the property of collaboration, which enables cooperation with other actors (Liedtke et al., 2012). Consequently, collaboration is also related to the use of participatory methods such as co-creation (Franz, 2015; Liedtke et al., 2012). In addition diverse sources indicated the characterization of sustainability. This means the innovations developed in Living Labs often have a tangible sustainability effect (Liedtke et al., 2012). The general duration of projects in Living Labs are planned on a medium- to long-term view (Kanstrup, 2017; Ley et al., 2015). Furthermore, products and services are tested in Living Labs in a way as if they were found to be safe and risk-free (Engels et al., 2019). The plot size for conducting the experiment is small to medium. For example, measured by the number of households, one research conducted in 8 households and another research conducted with 17 households (Ley et al., 2015). An overview of the characterization of Living Labs is provided in Table 3. In sum research in Living Labs is user-focused and conducted in a real-life environment (Dell'Erra and Landoni, 2014; Kanstrup, 2017; Geibler et al., 2013; Liedtke et al., 2015; Liedtke et al., 2012). Examples of existing Living Labs are Fraunhofer-InHaus-Zentrum in Duisburg[1], Germany and GovLabAustria[2].

**Table 3.** Overview of the characterization of Living Labs

| Typification | Source |
|---|---|
| Sustainability | Liedtke et al., 2012; Geibler et al., 2013; Leminen et al., 2016; Hossain et al., 2019 |
| Open Innovation | Liedtke et al., 2012; Hossain et al., 2019 |
| Collaboration | Liedtke et al., 2012 |
| Real-life environment | Kanstrup, 2017; Liedtke et al., 2012; Liedtke et al., 2015; Geibler et al., 2013; Bergvall-Kareborn et al., 2015; Franz, 2015; Franz et al., 2015; Hossain et al., 2019 |

---

[1] See https://www.inhaus.fraunhofer.de/
[2] See https://www.govlabaustria.gv.at/

| Medium to long-term orientation | Ley et al., 2015; Kanstrup, 2017 |
| User-focused | Kanstrup, 2017; Franz, 2015; Hossain et al., 2019 |
| Risk-free | Engels et al., 2019 |
| Products and Services | Franz, 2015; Zheng, 2016 |
| Participative | Franz, 2015 |
| Co-creation | Franz, 2015 |
| Real-world experiments | Kanstrup, 2017 |
| Medium to large area size | Ley et al., 2015; Følstad, 2008; Cardullo et al., 2018 |

## b. Real World Laboratories

In recent years, a number of definitions and characteristics for Real World Laboratories have been developed. These converge on important points, which makes it easy to facilitate an approach regarding a consensus of Real World Laboratories (Wagner and Grunwald, 2019). In this work, the following criteria are therefore considered necessary to define a Real World Laboratory (Schneidewind, 2014; Rose et al., 2019): 1) transdisciplinarity as a central research approach 2) (societal) learning processes and continuous reflection of the approach 3) long-term orientation, scalability, and transferable solutions to societal problems 4) real-world interventions (so-called real world experiments) 5) contribution to sustainability transformation. In the analysis of Real World Laboratories, one of the insights is that Real World Laboratories increasingly engage in transdisciplinary research, a type of research that is reflexive and self-reflexive, organizing research as a joint learning process between society and science (Rose et al., 2019; Schäpke et al., 2017). Additionally, in the collaboration of researchers and the participating stakeholders in Real World Laboratories, development is done with focus on sustainability (Engels et al., 2019). The analysis also showed that real-world problems have served as a starting point for Real World Laboratories (Engels et al., 2019). Real-world problems deal with issues that can be transferred from science to practice. These include, for example, strategies of building out an entire city or developing residential communities (Schneidewind, 2014; Schäpke et al., 2017). Also, social, political, and even sociotechnical issues are affected by real-world problems (Schäpke et al., 2017). Due to the complexities of the issues that Real World Laboratories address, the duration of research is therefore designed to be long-term. Stakeholders are involved in the research in a participatory manner. This can range from consultation, information up to co-creation (Engels et al., 2019). The area size in which Real World Laboratories experiment range from several households to entire neighborhoods or cities (Schäpke et al., 2017). Examples of existing Real World Laboratories according to the above definition are the project "Wohlstands-Transformation Wuppertal"[3] in Wuppertal, Germany and BaWü-Labs[4] in Germany. Table 4 provides an overview about their characteristics:

---

[3] See https://w-indikatoren.de/die-transformation/
[4] See www.reallabore-bw.de

**Table 4.** Overview of the characterization of Real World Laboratories

| Typification | Source |
|---|---|
| Transdisciplinary | Rose et al., 2019; Schäpke et al., 2017 |
| Sustainability | Rose et al., 2019; Schäpke et al., 2017 |
| Long-term orientation | Rose et al., 2019; Wirth et al., 2019 |
| Transferable problems | Rose et al., 2019 |
| Regulatory | Rose et al., 2019 |
| Real world problems | Rose et al., 2019 |
| Participative | Engels et al., 2019 |
| Large area size | Schneidewind, 2014; Schäpke et al., 2017 |

### c. Novel test beds

Test beds have existed before the creation of Living Labs and Real World Laboratories and have supported research for quite some time. Therefore, the term "novel" takes on a different meaning in relation to test beds. For example, they provide experimental environments to test products in a controlled setting or to observe experiments and analyse test data (Wanfeng et al., 2018). The application areas of testbeds are wide-ranging in this regard. They range from military test trials, testing of new products, to new test tracks for autonomous driving in the field of mobility. What can make test beds "novel research infrastructures" is their ability to integrate novel methods and technologies into the research infrastructure. Thus, with regard to test beds, those that exhibit the characteristics of a novel research infrastructure mentioned above are considered novel research infrastructures. Thus, along with Living Labs and Real World Laboratories, novel test beds have also emerged in recent years as an important approach to foster innovation across geographic regions and technical domains (Engels et al., 2019). Novel test bed projects for smart and sustainable cities, whether in China (Tianjin) or Abu Dhabi (Masdar City), experiment with combining innovation and urban living to enable both new forms of urbanity and new forms of innovation, often with the goal of becoming a model for other cities (Engels et al., 2019). Novel test beds generally aim to support innovation. Specifically, novel test beds can, on the one hand, raise awareness of the importance of innovation activities for competitiveness or, on the other hand, help overcome systemic failures in innovation. Moreover, with the help of the use of novel test beds, users can be involved in the innovation process (Ballon et al., 2005). In current research the term novel test bed is referred to very differently such as experience and application research centers, prototyping environments, field trials, co-development environments, user trials, and many more (Ballon et al., 2005). Despite the fact that no uniform terminology is used, the used terms generally have in common that a distinction is made between environments for testing and environments for design and development (Ballon et al., 2005). In novel test beds, newly developing, unfinished, and potentially risky technologies are adopted only on a trial basis because certain design issues regarding their various risks and safety aspects can only be resolved based on empirical usage data (Engels et al., 2019). Through these types of novel test beds, innovative technologies that still pose risks to society can be fully tested in a real-life environment (Engels et al., 2019). In novel test beds, the experiment's environment is not fixed. It is possible to experiment here in a real-life environment or in a laboratory-like environment. Moreover, this environment can be controlled, semi-controlled, or uncontrolled (Wanfeng et al., 2018; Engels et al., 2019; Ballon et al., 2005). Thus, the configuration of the environment for testing the innovation is diverse. Furthermore, the design remains open in terms of the structure of the novel test beds. This can be carried out in a closed model or in an open one. The latter makes it possible to experiment in cooperation with other actors and to get research results from a different point of view in order to conclude new insights. This also makes it possible to use participatory methods such as co-evaluation (Ballon et al., 2005). Since legal exemptions exist in testbeds with respect to experimenting with innovations, it is possible to introduce and experimentally adopt emerging, unfinished, and potentially risky technology in testbeds in a real-life environment, because certain design issues related to risk and safety can only be re-clarified based on empirical usage data (Rose et al., 2019). One example of an exceptional novel test bed is the city of Songdo in South Korea, which was built from scratch as a technologically highly integrated city.[5]

**Table 5.** Overview of the characterization of novel test beds

| Typification | Source |
|---|---|
| Real-life environment, laboratory-like environment, controlled, Semi-controlled | Wanfeng et al., 2018; Engels et al., 2019; Ballon et al., 2005 |
| Tests of risky innovations | Engels et al., 2019 |
| Open innovation model, closed innovation model | Wirth et al., 2019, |
| Participation | Wirth et al., 2019; Ballon et al., 2005 |
| Sustainability | Engels et al., 2019; Wirth et al., 2019; Ballon et al., 2005 |

### d. Differentiation of novel research infrastructures

The three research infrastructures presented are currently not differentiated uniformly in research (Ballon et al., 2005). For example, in (Fuglsang et al., 2021) Living Labs and Real World Laboratories are generalized and thus Real World Laboratories are also considered Living Labs. On the other hand (Schäpke et al., 2017) differentiate Living Labs and Real World Laboratories. Other works consider test beds and Living Labs separately and thus define them as a separate type of research infrastructure (Engels et al., 2019). The fact that no consensus exists regarding definitions makes it difficult to distinguish between these research infrastructures and to provide them with a scientific foundation. In this section, we propose an approach to distinguish these three novel research infrastructures.

Real World Laboratories and Living Labs overlap in some characteristics, for example, in terms of a focus on sustainability issues, the duration of research, the principle of open innovation, and the real-life environment. For this reason, terms such as Real World Laboratories, Living Labs, urban Living Labs, sustainable Living Labs, and even Smart Cities often lack consensus among different research efforts (Soeiro, 2021). This also mixes analyses of whole-urban transformations in Real World Laboratories with neighborhood approaches and household-based Living Labs (Liedtke et al., 2012; Ahmadi et al., 2020). For example, (Voytenko et al., 2016) assigns Urban Living Labs and Smart Cities to the Living Labs category while (Schneidewind, 2014; Schäpke et al., 2017) considers Urban Living Labs and Sustainable Living Labs as a separate Real World Laboratory apart from a Living Lab. On the other hand, research exists that considers Urban Living Labs separately from Living Labs (Soeiro, 2021). In order to distinguish urban Living Labs from Real World Laboratories in particular, (Schneidewind, 2014) has proposed a concept to differentiate between the two terms. Here, the level in terms of spatiality and scale is used as the main differentiator. The "household level" includes individual households or even blocks of flats in which "household-based intervention strategies" are used (Liedtke et al., 2015). The term Living Labs has become internationally established for this level (Kareborn and Stahlbrost 2009). Therefore, in this type of level, social interactions within a large number of households, which is associated with a complex network, can be difficult to study. The neighborhood level includes city neighborhoods or even districts. Here it is possible to observe effects of cultural identities and social diffusion processes (Schneidewind, 2014). It is at this scale that the term Real World Laboratory comes into play (Schäpke et al., 2017; Rose et al., 2019; Rogga et al., 2018). The findings from Real World Laboratories at this scale provide a higher degree of cross-city comparability. The third level is referred to by (Schneidewind, 2014) as the "city level." This encompasses city-wide processes. At this level, a large number of transformation processes play a relevant role in research, as only at this level the inclusion of effects of city-wide infrastructures is possible (Schneidewind, 2014). This distinction according to (Schneidewind, 2014) on three levels makes it possible to create a basis for the differentiation of the terms Living Labs and Real World Laboratories. Here, the difference between the two according to (Schneidewind, 2014) lies firstly in the area of the experiment to be observed. Experiments on a large-scale level, such as a city, in which cultural changes or even overall societal developments are investigated, call for Real World Laboratory approaches, while Living Lab approaches are suitable on a household level. Household level refers to households or blocks of flats in which new technologies or intervention strategies are studied in particular (Liedtke et al., 2015). Thus, the spatial environment can be used to distinguish between Living Labs and Real World Laboratories. Research work in Real World Laboratories usually covers larger areas, such as larger city districts or even entire cities. In contrast, research work in Living Labs also covers isolated households. However, this characteristic alone is not sufficient to distinguish Living Labs from Real World Laboratories. This is also shown by the research of Cardullo et al. (2018). In this paper, it is clarified that Living Labs in practical applications also include the size of a city. Further distinctions can be based on the time period. In this context, Living Lab research approaches are usually designed for the medium to long term, as the developed prototypes, products, or even services must be brought to a certain level of market maturity before they can be released for use (Følstad, 2008). Research in Real World Laboratories, on the other hand, is fundamentally long-term in nature and addresses complex problems; mostly in sociotechnical fields or in transformative sustainability research (Schäpke et al., 2017; Ferronato et al., 2019).

Inconsistencies can also be seen when considering Living Labs and test beds. Zhong et al. (2006) and Oliveira et al. (2006) presented Living Labs as a kind of test beds extended by the previous phases initialization and development. On the other hand, Ballon et al. (2005) call for Living Labs to be considered separately from test beds. This work agrees with Ballon et al. (2005).

In order to differentiate the three research infrastructure models, in this work an overview (see Table 6) is provided, which classifies them according to the dimensions "Thematic Focus", "Spatial Focus", "Time Frame", "Environment" and „Research methods".

**Table 6.** Differentiation of Living Labs, Real World Laboratories and novel test beds

| Dimension | Living Lab | Real World Laboratory | Novel test bed |
|---|---|---|---|
| **Thematic Focus** | - Products and services (Zheng, 2016; Liedtke et al., 2012)<br>- Sustainability (Liedtke et al., 2015; Leminen et al., 2016; Geibler et al., 2013) | - Transferable problems (Schäpke et al., 2017; Schneidewind, 2014; Rose et al., 2019)<br>- Social and political problems (Schäpke et al., 2017)<br>- Sustainability (Schäpke et al., 2017)<br>- Socio-technical problems (Schäpke et al., 2017) | - Detection of risks of already developed products (Zheng, 2016; Engels et al., 2019)<br>- Sustainability (Engels et al., 2019; Wirth et al., 2019; Ballon et al., 2005) |
| **Spatial Focus** | - Few to various households (Ley et al., 2015)<br>- Partly also large number of households (Følstad, 2008)<br>- City (Cardullo et al., 2018) | - International, urban district, city (Schneidewind, 2014) | |

| Time frame | - Medium-term (Ferronato et al., 2019; Schäpke et al., 2017)<br>- Long-term (Kanstrup, 2017) | - Long-term structures and subsequent projects (Wirth et al., 2019; Schäpke et al., 2017) | |
|---|---|---|---|
| Environment | - Real-life environment (Kanstrup, 2017; Bergvall-Kareborn et al., 2015; Franz, 2015; Vale et al., 2018)<br>- Participative, external control (Schäpke et al., 2017; Franz, 2015; T. Vale et al., 2018)<br>- controlled (Nyström et al., 2014) | - Real-life environment<br>- Participative respectively low controllability (Rose et al., 2019; Schäpke et al., 2017) | - Real-world or laboratory-like environment (Engels et al., 2019; Ballon et al., 2005)<br>- Controlled (Engels et al., 2019) |
| Research methods | - Participative (Gansl, 2020; Ley et al. 2015; Ballon und Schuurman, 2015)<br>- Open innovation model (Liedtke et al., 2012; Liedtke et al., 2015) | - Transdisciplinary (Rose et al., 2019; Schäpke et al., 2017)<br>- Transformative (Schäpke et al., 2017)<br>- Open Innovation (Schäpke et al., 2017; Rose et al., 2019) | - Open or closed innovation model (Wirth et al., 2019) |

The "Thematic Focus" dimension summarizes different (thematic) research objects with which the research infrastructures are fundamentally concerned. The comparison shows that living labs primarily address the generation of innovations in the area of product development (Liedtke et al., 2012). In contrast, Real World Laboratories deal with more complex societal problems. Nevertheless, in both research infrastructures, research fields can be found in which both are active; for example, in sustainability topics. In contrast, in the case of test fields, no explicit topic is specified or described in the sources analysed. Although sustainability issues can also be researched with test beds, it is not explicitly mentioned as a focus as in the case of Real World Laboratories or Living Labs. (Engels et al., 2019) explicitly state in their research on test beds that compared to Living Labs, test beds have legal leeway to implement and test innovations. This allows, for example, risky innovations to be tested and their data subsequently validated to minimize these risks. The „Spatial Focus" dimension describes the research area of the experiment. Here, the difference between Living Labs and Real World Laboratory can be seen, as living labs can also conduct research on smaller areas, such as a few households. Research conducted by Real World Laboratories, on the other hand, is predominantly conducted on larger areas such as neighborhoods or cities (Rose et al., 2019; Schneidewind, 2014). Regarding the spatial area of test beds, the analysis could not find any results from the sources of this work. In the dimension of "Time frame", a difference can be identified between Real World Laboratories and Living Labs. While research from Living Labs can also be long-term, these are generally medium-term. In contrast, research work in Real World Laboratory runs over several years. No explicit information can be found on the duration of research work in novel test beds.

## 5. Conclusion

In this paper it was recognized that currently used definitions of the novel research infrastructures described in this work – namely Living Labs, Real World Laboratories and novel test beds - do not offer a uniform consensus. The characteristics and structures of such research infrastructures vary greatly in practice - even from Living Lab to Living Lab - as there is almost no established framework regarding their setup. This work provides an approach to distinguish the three identified novel research infrastructures. Based on that, this approach can be used in the future to investigate further distinctive properties and theoretical foundations addressing an urgent need to understand the evolution of the key facets of Living Labs and similar research infrastructures, such as the characteristics and outcomes (Hossain et al.; 2019). This, in turn, may then also allow for more precise decision making as to which research infrastructure is suitable for which research question. A follow-up question is in particular how a new cooperative research infrastructure should be set up which - like the KARL project (cf Alpers 2022), for example - explicitly not only wants to enable cooperative innovations but also pursues a human-centered approach, e.g., considering work quality.

## Acknowledgements

The results presented here were partly developed within the project 'Competence Centre KARL – Artificial Intelligence for Work and Learning in the Karlsruhe Region' (https://kompetenzzentrum-karl.de). This research and development project is funded by the German Federal Ministry of Education and Research (BMBF) within the program 'The Future of Value Creation – Research on Production, Services and Work' (funding number 02L19C253) and is managed by the Project Management Agency Karlsruhe (PTKA). The authors are solely responsible for the content of this publication.

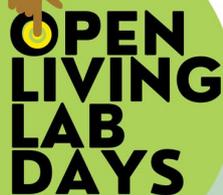

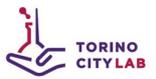
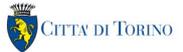
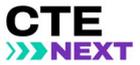
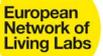